\makeatletter
\def\input@path{ {.} {tex} {fig} {tblr} {sty} }
\makeatother
\documentclass[%
	a4paper,UKenglish,numberwithinsect,
	cleveref,autoref,thm-restate,subcaption,
	nameinlink,colorlinks,pdfusetitle,
	svgnames,
	inline,
]{lipics-v2021}
\usepackage{fix-lipics-v2021}

\usepackage{amssymb}
\usepackage{ragged2e}
\usepackage{enumitem}
\usepackage{multicol}
\usepackage{tabularray}
\usepackage{standalone}

\usepackage[T1]{fontenc}
\usepackage{lmodern}
\usepackage{bold-extra}

\usepackage{macros}
\usepackage{localtodo}
\usetikzlibrary{gb-loops}

\nolinenumbers
\hideLIPIcs
\def\subjclassHeading{}
\def\keywordsHeading{\vspace{-2\baselineskip}}

\title{Polynomial Complementation of Nondeterministic $2$\=/Way Finite Automata by $1$\=/Limited Automata}

\author{Bruno Guillon}{Université Clermont Auvergne, Clermont Auvergne INP, LIMOS, CNRS, F\=/63000 Clermont-Ferrand, France}{bruno.guillon@uca.fr}{0000-0003-1630-3404}{}
\author{Luca Prigioniero}{Department of Computer Science, Loughborough University, Epinal Way, Loughborough LE11 3TU, UK}{l.prigioniero@lboro.ac.uk}{0000-0001-7163-4965}{}
\author{Javad Taheri}{Université Clermont Auvergne, Clermont Auvergne INP, LIMOS, CNRS, F\=/63000 Clermont-Ferrand, France}{javad.taheri369@gmail.com}{0009-0007-5106-2243}{}

\authorrunning{B. Guillon, L. Prigioniero, and J. Taheri}
\Copyright{Bruno Guillon, Luca Prigioniero, and Javad Taheri}
\date{April 2025}

\begin{document}
\maketitle

\begin{abstract}
	We prove that,
	paying a polynomial increase in size only,
	every unrestricted two-way nondeterministic finite automaton (\twnfa)
	can be complemented by a 1\-/limited automaton (\la),
	a nondeterministic extension of \twnfas
	still characterizing regular languages.
	The resulting machine is actually a restricted form of~\las{}
	--- known as \twnfas with common guess ---
	and is self-verifying.
	A corollary of our construction is that a single exponential is necessary and sufficient
	for complementing \las.
\end{abstract}

\section{Introduction}
\label{sec:intro}

The study of the resources used by computational models
is a central topic in automata theory.
One classical problem in this area is to determine the cost of applying operations between languages
(\eg, union, intersection, concatenation, Kleene star, \etc).
Here,
the cost is defined as the increase in size of the resulting (or \emph{target}) devices
after applying the operation to the languages recognized by the original (or \emph{source}) machines.

In this paper, we focus on the cost of the complementation of regular languages.
This operation is usually cheap (\ie, costs at most polynomial)
when dealing with deterministic devices,
while it is often expensive (\ie, at least exponential)
for nondeterministic devices;
see~\cref{tab:costs}.
The reason for this separation has been understood for a long time
and originates from the nature of nondeterminism,
as illustrated by the case of classical
(\emph{one-way})
nondeterministic finite automata (\ownfas).
Indeed, the semantics of such a device
is that a word is accepted as long as \emph{there exists} a computational path
leading to an accepting state.
Therefore, in order to acknowledge that a word does not belong to the recognized language,
one should somehow check that \emph{every computational path} leads to a non-accepting state,%
\footnote{\label{fn:complete automata}For ease of discussion, we admit here that the automata are complete, that is, no computational path gets stuck in the middle of the input.}
a semantic which is hardly captured by nondeterminism.
This issue does not exist for one-way deterministic finite automata (\owdfas),
because they admit a unique computational path on each input,
and thus existential and universal quantification on computational paths collapse.
Indeed,
it is folklore that
exchanging accepting and non-accepting states of a \owdfa,
while keeping the rest of the structure (initial state and transitions) unchanged,
yields a \owdfa recognizing the complement of the language.\footref{fn:complete automata}
On the other hand,
it is well known
that the transformation of a \ownfa into another one recognizing the complement of the language
may cost as much as determinizing it and then complementing it (as explained above) in the worst case~\cite{SS78,Bir92,Jir05}.
\begin{table}[tb]
	\hfill
	\begin{tblr}[t]{colspec={Q[4em]Q[7.5em]},hlines,vlines,row{1}={font=\strongsty},column{1}={c}}%
		model		&	cost										\\
		\owdfa	&	trivial									\\
		\twdfa	&	linear~\cite{GMP07}			\\
		\dla		&	poly~\cite{GP19}\newline
	\end{tblr}
	\hfill\hfill
	\begin{tblr}[t]{colspec={Q[4em]Q[15em]},hlines,vlines,row{1}={font=\strongsty},column{1}={c}}
		model		&	 cost																			\\
		\ownfa	&	 exp~\cite{SS78}														\\
		\twnfa	&	 ??? (related to SS78, \via~\cite{GMP07})	\\
		\la			&	 exp (lower bound in~\cite{PP14,GPT25}, and\newline\vbox{}\hfill upper bound in \cref{cor:complement 1-la})
	\end{tblr}
	\hfill\vbox{}
	\caption{%
		Tight cost orders for the complementation
		on different models of finite automata.
		Here,
		the target device is the same as the source device,
		and it is indicated in the first column.
		``SS78'' stands for ``the Sakoda and Sipser conjecture''~\cite{SS78}.
		It can be observed that the transformation is cheap (\ie, at most polynomial) for deterministic devices,
		and expensive (\ie, exponential) or unknown for nondeterministic ones.%
	}
	\label{tab:costs}
	\vskip-1.5\baselineskip
\end{table}

Yet, the corresponding question remains unsolved for other regular language recognizers,
and in particular when dealing with \emph{two-way finite automata},
an extension of finite automata allowing the machine to move its head both back and forth,
and which still characterizes regular languages.
Indeed, although complementing two-way deterministic automata (\twdfas)
has been non-trivially shown to cost linear only~\cite{GMP07},
the cost for complementing their nondeterministic counterparts (\twnfas)
is still unknown in the general case.%
Worse still, the best-known upper bound is exponential
and is obtained by transforming the source~\twnfa into an equivalent~\owdfa%
.
That is, neither two-wayness nor nondeterminism
are exploited for complementing arbitrary~\twnfas.
Indeed, also the cost for determinizing \twnfas is a longstanding open question
known as \emph{``the Sakoda and Sipser problem''}~\cite{SS78} (see \eg~\cite{Pig13} for a survey).
As shown in~\cite{GMP07},
the two problems are related
\via the linear-cost complementation of~\twdfas.
On the one hand, finding an exponential (or super-polynomial) lower bound for complementing~\twnfas
would imply a similar lower bound for determinizing them.
On the other hand, finding a polynomial (or sub-exponential) upper bound for determinizing~\twnfas
would imply a similar upper bound for complementing them.
It is worth noting
that a polynomial-cost complementation of \twnfas
has been obtained in some particular cases,
\eg, in the \emph{unary}
setting~\cite{GMP07},
or when the \twnfa makes a restricted use of nondeterminism,
known as \emph{outer-nondeterminism}~\cite{GGP14}.%

In this paper,
we study the cost of complementing two-way finite automata following a different approach:
Instead of using the same model as source and target devices,
we relax the target machines to have some rewriting capabilities.
In particular,
we use as a target device a machine called \emph{1\=/limited automaton} (\la),
which is an extension of \twnfas
that can rewrite the contents of the tape
\emph{only} during the first visit to each cell.
This model is not more powerful than \twnfas~\cite{WW86},
\ie, it recognizes regular languages only.
However, there are cases where it can represent languages more succinctly than \twnfas
(for a recent survey on this model, see~\cite{Pig19}).
This approach has already been used to provide succinct representations of
operations that have exponential cost when both source and target machines are \owdfas
(Kleene star, reversal and concatenation),
but can be done at a polynomial cost when the target machine is a \emph{deterministic \la} (\dla)~\cite{PPS24}.

\begin{figure}
	\centering
	\def\qmark{\strong{?}\xspace}
\begin{tikzpicture}[x=64mm,y=25mm]
	\path[every node/.style={minimum width=3.125em,inner sep=2pt,draw=black,thin,shape=rectangle}]
	(0,0)	node	(1la)		{\la}
	(0,1)	node	(2nfa)	{\twnfa}
	(1,0)	node	(d1la)	{\dla}
	(1,1)	node	(2dfa)	{\twdfa}
	;
	\path[every node/.style={inner sep=1pt}, every edge/.style={-latex,draw=blue,thick}]
	(2nfa)	edge						node[above]																											{\qmark~\cite[SS78 Prob.]{SS78}}	(2dfa)
	(1la)		edge						node[below]																											{$\leq$\qmark~\cite[Prob.2]{Pig19}}	node[above] {$\geq$exp~\cite{PP14,GPT25}} (d1la)
	(2nfa)	edge						node[above,sloped]																							{\qmark~\cite[Prob.4]{Pig19}}	(d1la)
	;
	\path[every edge/.style={-latex,draw=red,dashed,thick}]
	(2nfa)	edge[loop west]	node[left]																											{\qmark~\cite{GMP07}}	(2nfa)
	(1la)		edge[loop west]	node[left,text width=21.5mm,align=right]												{$\geq$exp~\cite{PP14,GPT25}\linebreak$\leq$exp \xref[Cor.]{cor:complement 1-la}}	(1la)
	(2dfa)	edge[loop east]	node[right]																											{lin~\cite{GMP07}}	(2dfa)
	(d1la)	edge[loop east]	node[right]																											{poly~\cite{GP19}}	(d1la)
	(2nfa)	edge						node[left,text width=14.5mm,align=right,xshift=-1mm,yshift=1mm]	{poly \xref[Thm.]{thm:2nfa to sv1-la}}	(1la)
	;
\end{tikzpicture}
	\caption{%
		Size cost orders for various transformations discussed in introduction.
		Plain blue arrows mean conversions of sources into equivalent targets,
		while dashed red arrows mean complementations of sources with targets.
		Question marks indicates the open problems,
		including~``the Sakoda and Sipser conjecture'' denoted as~``SS78''~\cite{SS78}.%
	}
	\label{fig:map}
\end{figure}
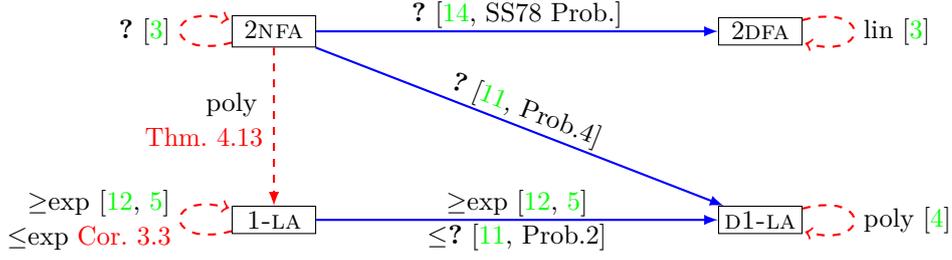

In the survey~\cite{Pig19}, the author identifies some problems regarding the descriptional complexity of \las.
In particular, the question of the cost of the conversion of \twnfas into equivalent \dlas~(Problem~4),
as well as those of the cost of determinizing \las (Problem~2) are raised.
Just as complementing \twnfas relates to the Sakoda and Sipser problem,
complementing \twnfas with \las relates to these problems;
see \cref{fig:map}.

\subparagraph{Our results.}
We show polynomial simulations of \ownfas and \twnfas by \emph{self-verifying} \las (\cref{thm:1nfa to sv1-la,thm:2nfa to sv1-la}),
the property of being self-verifying meaning that the devices are able to recognize both the language and its complement
(the formal definition is given in \cref{sec:prelim}).
In both constructions,
the resulting devices are \las of particular form,
known as \emph{\twnfas with common guess} (\twnfacgs),
in which the rewriting of the tape
is made during an initial nondeterministic memoryless traversal of the input
-- see~\cite{GP19,GPT25} for details and results on this model.
Although a polynomial simulation of \ownfas by self-verifying \las is implied by \cref{thm:2nfa to sv1-la},
this particular case is treated in \cref{thm:1nfa to sv1-la},
which presents a specific construction for this case
that is simpler, cheaper, and serves as a preparatory step for the more technical second construction.
Also,
because every \la can be converted into a \ownfa paying a single exponential~\cite{PP14},
a consequence of \cref{thm:1nfa to sv1-la}
is a single exponential upper bound for the cost of the complementation of~\las (\cref{cor:complement 1-la}).
This improves the best-known upper bound for the transformation,
which used the simulation of \las by \owdfas at a doubly-exponential cost.
Since an exponential lower bound is known for this transformation~\cite{GPT25},
the cost order is tight.
\Cref{fig:map} summarizes our results and their connection to open questions and some related results.

\subparagraph{Outline.}
The paper is organized as follows.
\Cref{sec:prelim} gathers the definitions and notations used throughout the subsequent sections.
In \cref{sec:1nfa} we present the conversion of \ownfas into self-verifying \twnfacgs,
and its consequence on the cost of the complementation of \las.
In \cref{sec:2nfa} we develop the conversion of \twnfas into self-verifying \twnfacgs.
A brief conclusion is given in \cref{sec:conclusion}.

\section{Preliminaries}
\label{sec:prelim}

In this section,
we recall some fundamental definitions and notations used throughout the paper.
We assume that the reader is familiar with basic concepts from formal languages and automata theory
(see, \eg,~\cite{HU79}).

For a set~$S$,
$\defem{\card{S}}$ denotes its cardinality
and~\defem{$\powerset{S}$} denotes its powerset.
For~$n\in\Nat$,
\defem{$\interv{n}$} denotes the set consisting of the first~$n$ natural numbers (including~$0$),
namely~$\interv{n}=\set{0,\ldots,n-1}$;
in particular~$\interv{0}=\eS$ and~$\interv{1}=\set{0}$.
Given an alphabet~$\ialphab$,
the set of strings over~$\ialphab$
is denoted by~\defem{$\ialphab*$}.
It includes the empty string denoted by~\defem{\ew}.
The length of a word~$w\in\ialphab*$ is denoted by~\defem{$\lvert w\rvert$},
and the set of strings over~$\ialphab$ of length~$i$ is denoted by~\defem{$\ialphab^i$}.
The number of occurrences of a symbol $\sigma\in\ialphab$ in $w$ is denoted by $\length{w}_\sigma$.
The positions (or indices) of symbols within a string~$w$
are indexed from~$0$ to~$\length{w}-1$.
We use the notation~\defem{$w[i]$} to indicate the symbol at position~$i\in\interv{\length{w}}$ of~$w$
(\ie, the~$(i+1)$\=/th symbol of~$w$, \eg, $w[0]$ is the first symbol of~$w$)
and~\defem{$w[i,j]$} for the factor of~$w$ from index~$i$ to index~$j$ included,
$0\leq i, j<\length{w}$.
If~$i>j$, we set~$w[i,j]=\ew$ by convention.
Hence, the length of~$w[i,j]$ is~$0$ if~$i>j$ and~$j-i+1$ otherwise.

\begin{definition}
	\label{def:2nfa}%
	A \defem{two-way nondeterministic finite automaton} (\defem{\twnfa})~${\mach{A}}$
	is a tuple ${\struct{Q,\ialphab,\delta,\qstart,\qfin}}$,
	where~$Q$ is the finite set of \defem{states},
	$\ialphab$ is the \defem{input alphabet},
	$\qstart\in Q$ is the \defem{initial state},
	$\qfin$ is the \defem{final state},%
	\footnote{%
		\label{fn:unique final state}%
		Notice that, for convenience, we shall assume that \twnfas have a unique final state.
		This is not a limitation, since each $n$\=/state \twnfa with many final states
		can easily be simulated by a ${(n+1)}$\=/state \twnfa with, as in our definition, a single final state.
		(We do not consider one-way deterministic finite automata, for which this restriction makes a difference.)
		The same comment applies on initial states.%
	}
	and~$\delta:Q\times\ialphab|\rightarrow 2^{Q\times\set{\lmove,\rmove}}$
	is a nondeterministic \defem{transition function}
	with ${\defem{\ialphab|}=\ialphab\cup\set{\lend,\rend}}$,
	where $\lend,\rend\notin\ialphab$
	are two special symbols called \defem{the left} and \defem{the right endmarker}, respectively.%
	\label{def:2nfa}%
\end{definition}
We shall often assume~$\Q=\interv{n}$ for~$n=\card{\Q}$,
so that~$\Q$ is ordered,
with~$-1\notin\Q$ as a minimum.

In \twnfas,
the input is written on the tape surrounded by the two endmarkers,
the left endmarker being at tape position zero.
Hence, on input~$w$, the right endmarker is at position~${\length{w}+1}$.
In one move,
\mach{A} reads an input symbol,
changes its state,
and moves the head one position backward or forward
depending on whether~$\delta$ returns~$\lmove$ (a \defem{left move})
or~$\rmove$ (a \defem{right move}),
respectively.
Furthermore,
the head cannot pass the endmarkers.
The machine accepts the input
if there exists a \emph{computational path} which
starts from the initial state~\qstart
with the head on the cell at position~$1$
(\ie, scanning the first letter of~$w$ if~$w\neq\emptyword$ and scanning~$\rend$ otherwise)
and eventually halts in the final state~$\qfin$ with the head scanning the right endmarker.
The language accepted by~\mach{A} is denoted by~\defem{$\langof{\mach{A}}$}.

A \twnfa is \defem{one-way} (\ownfa) if its head can never move left, \ie, if no transition returns~$\lmove$.
The transition function of a \ownfa
is seen as a function~$\delta:\Q\times\ialphab\to\powerset{\Q}$
(that is, the endmarkers and the instruction for head direction are irrelevant).

The above models are all read-only machines.
We now extend \twnfas with a limited write ability.
A \defem{$1$-limited automaton} (\la, for short) \mach{A}
is a~\twnfas over~$\ialphab$
that, on every step on symbol~$\sigma$,
rewrites it with some symbol~$\tau\notin\ialphab$ from a \emph{work alphabet}~$\walphab\supset\ialphab$
if the contents of the cell has not been already rewritten,
namely if~$\sigma\in\ialphab$.
In other words,
\mach{A} can modify the contents of a cell
only when it visits the cell for the first time and the cell does not contain an endmarker.
Acceptance for~\las is defined exactly as for \twnfas,
and the language accepted by a given \la~$\mach{A}$ is denoted by~\defem{$\langof{\mach{A}}$}.

\subparagraph{Common guess.}
In this paper,
we use as target devices particular cases of \las,
whose computations are somehow split into two phases.
In the first phase, 
using one state only,
these machines make a single one-way pass over the input
during which they nondeterministically rewrite (or \emph{annotate}) every tape cell symbol,
and then move the head back to the left endmarker.
In the second phase,
they perform
read-only two-way computations on the annotated word.
In other words,
this model can be seen as an extension of \twnfas
with the ability (called \emph{common guess}) to initially annotate the input word~$w\in\ialphab*$
using some annotation symbols from a fixed alphabet~$\Gamma$,
to which we refer as the \defem{annotation alphabet}.
The annotated word resulting from this initial phase
is a word over the product alphabet~$\ialphab\times\Gamma$.
It is nondeterministically chosen among all the words~$v\in{(\ialphab\times\Gamma)}^*$
such that~$\pi_1(v)=w$ (implying~$\lvert v \rvert =\lvert w\rvert$),
where~\defem{$\pi_1$} is the natural projection of~${(\ialphab\times\Gamma)}^*$ onto~$\ialphab*$.
We shall also use the notation~\defem{$\pi_2$} for the projection of~${(\ialphab\times\Gamma)}^*$ onto~$\Gamma^*$,
and the convention~$\pi_1(\lend)=\pi_2(\lend)=\lend$ and~$\pi_1(\rend)=\pi_2(\rend)=\rend$.
\begin{definition}
	\label{def:2nfa+cg}
	A {\twnfa} \emph{with common guess} (\defem{\twnfacg})
	is a triplet ${\mach{M}=\struct{\mach{A},\ialphab,\Gamma}}$
	where~$\ialphab$ is the input alphabet,
	$\Gamma$ is the \defem{annotation alphabet},
	and~\mach{A} is a \twnfa over the product alphabet~$\ialphab\times\Gamma$.
\end{definition}
The language recognized by~$\mach{M}$ is:
\begin{equation*}
	\defem{\langof{\mach{M}}}=\set{w\in\ialphab*\mid \exists v\in{(\ialphab\times\Gamma)}^*,\: v\in\langof{\mach{A}}\mathinner{\text{ and }}\pi_1(v)=w}=\pi_1(\langof{\mach{A}})
	\text.
\end{equation*}

In~\cite{GP19}, the authors showed that a polynomial increase in size
is always sufficient for the conversion of \la into \twnfacg.
Conversely, every $n$\=/state \twnfacg can be converted into an $(n+1)$\=/state \la.
For a recent discussion on this model,
we refer the reader to~\cite{GPT25}.

\subparagraph*{Configurations and computations.}
Given one of the devices~\mach{M} under consideration,
a \defem{configuration} is represented
as a string $\config{\lend x}{p}{y \rend}$,
meaning that~$p$ is the current state,
$xy\in\lend\talphab^*\rend$
is the contents of the tape
(%
	here~$\talphab$ denotes the \emph{tape alphabet}:
	$\ialphab$, $\walphab$, or $\ialphab\times\cgalphab$,
	depending on the model under consideration%
)
and the head is scanning the first symbol of~$y\rend$.
Hence, the \defem{initial configuration} over input~$w$
is~$\config{\lend}{\qstart}{w\rend}$,
and an \defem{accepting configuration} is a configuration of the form~$\config{\lend x}{\qfin}{\rend}$.
The transition relation between configurations of~\mach{M}
is denoted by~\defem{$\yield[M]$},
and its reflexive-transitive closure
by~\defem{$\yield*[M]$}.
A \defem{computational path} of~$\mach{M}$
is a sequence of configurations~$c_0,\ldots,c_r$
such that~$c_i\yield[M] c_{i+1}$ for each~$i<r$.
It is \defem{initial} (\resp \defem{accepting})
if~$c_0$ is initial (\resp~$c_r$ is accepting).
In order to emphasize the locality of some computational paths,
we also represent
\defem{partial configurations}
as~\config{u}{p}{v},
where~$p$ is the current state
and~$uv\in\set{\emptyword,\lend}\Delta^*\set{\emptyword,\rend}$
is a factor of the tape content.
The relations~$\yield[M]$ and~$\yield*[M]$,
whence the notion of computational path,
naturally extend onto partial configurations.

\subparagraph{Self-verifying.}
A nondeterministic state machine~$\mach{M}$ is said to be \defem{self-verifying}
if it has two different final states,
one \emph{accepting} and one \emph{rejecting},
and the set of words recognized by~\mach{M} is partitioned
into those admitting an \emph{accepting computational path} (\ie, ending in the accepting state),
and those admitting a \emph{rejecting computational path} (\ie, ending in the rejecting state).
In particular,
no word admits both an accepting and a rejecting computational path.
Computational paths that are neither accepting nor rejecting are called \emph{aborted}
(\ie, ending in a non-final state).

The set of words admitting an accepting computational path is denoted as~\defem{$\langof{M}$},
and the set of words admitting a rejecting computational path is its complement.

\subparagraph{Size of models.}
For each model under consideration,
we evaluate its size as the total number of symbols
used to describe it.
Hence, under standard representation and denoting by~$\ialphab$ the input alphabet,
the \defem{size} of an $n$\=/state \twnfa is~$\bigO{n^2\card{\ialphab}}$,
that of an $n$\=/state \la with work alphabet~$\walphab\supset\ialphab$ is~$\bigO{n^2\card{\walphab}^2}$,
and that of an $n$\=/state \twnfacg with annotation alphabet~$\cgalphab$ is~$\bigO{n^2\card{\ialphab} \cdot \card{\cgalphab}}$.
In our work, we generally consider~$\card{\ialphab}$ as a constant.

\section{Complementing \ownfas with \twnfacgs}
\label{sec:1nfa}

In this section, we present a construction
that transforms an arbitrary $n$\=/state \ownfa
into an equivalent self-verifying \twnfacg
with polynomially many states and~$2$ annotation symbols.
As in~\cite{GMP07,GGP14},
our simulation is based on the \emph{inductive counting} technique.

Let $\mach{A}=\struct{\Q,\ialphab,\delta,\qstart,\qfin}$ be an $n$\=/state \ownfa.
We assume~$\Q=\interv{n}$.
For~$w\in\ialphab*$, we define~\defem{${\QX[A]{w}}$}
to be the set of states the automaton~$\mach{A}$ reaches after reading~$w$,
\ie,
\(
	\QX[A]{w}	=\delta(\qstart,w)
	\text.
\)
Working on~$w$, a \twnfa can simulate~$\mach{A}$ several times,
bringing the head back to the left endmarker before each simulation.
It may thus find several states belonging to~$\QX[A]{w}$.
Suppose that the number~$m$ of states reached after reading~$w$ is given,
\ie,~$m=\card{\QX[A]{w}}$.
Then, by simulating~$m$ times~$\mach{A}$ on~$w$,
and controlling at each iteration but the first
that the state reached at the end of the simulation is larger than the preceding one,
a \twnfa can enumerate all the~$m$ states of~$\QX[A]{w}$.
This process is described in \cref{proc:enumQX}, 
where~\nsimulA
stands for the simulation of~$\mach{A}$ from the initial configuration.%
\footnote{%
	\label{fn:yield}%
	In enumeration algorithms,
	the outputs are not returned at the end of the execution,
	but yielded as soon as they are found.
	This allows an outer calling procedure to treat them one-by-one
	(see, \eg, \cref{proc:memberQX,proc:countnextQX}).
	In \cref{proc:enumQX}, this yielding is indicated with the \KwYield keyword
	on \cref{l:enumQX/yield}.%
}
\begin{figure}[tb]
	\begin{multicols}{2}
		\begin{procedure}[H]%
			\caption{%
				()\space%
				\enumQX{$m$}
				\label{proc:enumQX}
			}
			$\var{\q_{prev}}\gets-1$\;
			\For{$i\gets 1$ \KwTo~$m$}{%
				$\var{\q_{next}}\gets\nsimulA{}$\;
				\lIf{$\var{\q_{next}}\leq\var{\q_{prev}}$}{\Abort}
				$\var{\q_{prev}}\gets\var{\q_{next}}$\;
				\KwYield $\var{\q_{next}}$
				\label{l:enumQX/yield}
				\;
			}
		\end{procedure}
		\begin{procedure}[H]%
			\caption{%
				()\space%
				\checkAnnot{$\var{m}$}
				\label{proc:checkAnnot}
			}
			\tcp{starting from last position of~$x_i$}
			$\var{m}\gets\var{m}-\length{\pi_2(x_i)}_1$\tcp*[f]{move $n-1$ times leftward}\;
			\lIf{$\var{m}\neq0$}{\Abort}
			$\var{m}\gets\var{m}+\length{\pi_2(x_i)}_1$\tcp*[f]{move $n-1$ times rightward}\;
			\ForEach{$\p\in\enumQX{$m$}$}{%
				move~$n-\p-1$ times leftward\;
				\lIf{$\pi_2(\read{})\neq1$}{\Abort}
				move~$n-\p-1$ times rightward\;
			}
		\end{procedure}
		\begin{procedure}[H]
			\caption{%
				()\space%
				\memberQX{$\q_t$, $m$}
				\label{proc:memberQX}
			}
			\ForEach{$\var{\q}\in\enumQX{$m$}$}{%
				\lIf{$\var{\q}=\q_t$}{\Return \True}
			}
			\Return \False\;
		\end{procedure}
		\begin{procedure}[H]
			\caption{%
				()\space%
				\countnextQX{$m$}
				\label{proc:countnextQX}
			}
			$\var{m_{next}}\gets 0$\;
			\ForEach{$\var{\p}\in\Q$}{%
				\ForEach{$\var{\r}\in\enumQX{$m$}$}{%
					\If{$\var{\p}\in\delta(\var{\r},\read{})$}{
						$\var{m_{next}\gets\var{m_{next}}+1}$\;
						\Break\;
					}
				}
			}
			\Return $\var{m_{next}}$\;
		\end{procedure}
	\end{multicols}
	\vskip-1.75\baselineskip
\end{figure}
Hence,
according to whether one or none of them is the accepting state,
the \twnfa recognizes whether~$w$ belongs to~$\langof{\mach{A}}$ or to its complement;
see \cref{proc:memberQX} in which the accepting state is passed as a parameter~$\q_t$.
Remarkably,
a \twnfa~$\mach{B}_{m}$ with~$\bigO{n^2m}\subseteq\bigO{n^3}$ states
can implement this process.
Indeed, such a \twnfa only has to simultaneously store in its finite control:
the index~$i\leq m$ of the iteration, the previously found state ($\var{\q_{prev}}$), and the state in the current simulation ($\var{\q_{next}}$).
One strategy to detect whether an input~$w$ belongs to the complement of~$\langof{A}$
would therefore be to first compute~$m=\card{\QX[A]{w}}$,
and then run the \twnfa~$\mach{B}_{m}$.
Yet, computing~$\card{\QX[A]{w}}$ is not an easy task.
\medbreak

Suppose now that~$\card{\QX[A]{w}}$ is unknown,
but that there exists a nondeterministic procedure \nsimulA,
which, starting from some position~$\length{u}$,
where~$u$ is a prefix of the input~$w$,
eventually halts at the same position returning a state~$\q$
\ifof~$\q\in\QX[A]{u}$.
Then a \twnfa equipped with \nsimulA
may inductively compute~$\card{\QX[A]{u}}$ for each successive prefix~$u$ of~$w$ as follows.
Initially, $\card{\QX[A]{\ew}}=1$ since~$\QX[A]{\ew}=\set{\qstart}$ by definition.
Let~$u\sigma$ be a prefix of~$w$.
In order to compute~$\card{\QX[A]{u\sigma}}$ from~$\card{\QX[A]{u}}$,
the automaton tests for each state~$\p$ whether it belongs to~$\QX[A]{u\sigma}$,
and counts those for which the answer is positive.
This process is described in \cref{proc:countnextQX},
in which~$\sigma$ is read from the tape (indicated as~$\read{}$).
Testing whether a given state~$\p$ belongs to~$\QX[A]{u\sigma}$ is based on the following basic observation:
\begin{align}
	\label{eq:QX_usigma from QX_u}
	\p&\in\QX[A]{u\sigma}
	&	\iff	&&
	\exists\r\in\QX[A]{u}
	\mit*{such that}
	\p\in\delta(\r,\sigma)
	\text.
\end{align}
Using the knowledge of~$m=\card{\QX[A]{u}}$ and \nsimulA,
the automaton enumerates the~$m$ distinct states belonging to~$\QX[A]{u}$
in ascending order as explained before (\ie, using \cref{proc:enumQX}).
As soon as it finds one from which a transition on~$\sigma$ allows to enter~$\p$,
it breaks the loop as~$\p\in\QX[A]{u\sigma}$ has been witnessed.
If otherwise the~$m$ states are correctly found
but none of them has an outgoing transition to~$\p$ on~$\sigma$,
a witness of~$\p\notin\QX[A]{u\sigma}$ has been obtained.
Once the number of states in~$\QX[A]{u\sigma}$ has been computed,
the automaton forgets those of~$\QX[A]{u}$,
moves its head one cell to the right,
and proceeds with the next iteration of the induction.
\medbreak

However, implementing \nsimulA with a \twnfa is challenging.
Indeed, since there is an unbounded number of prefixes of inputs,
it is not possible to bring the head back to its initial position
(in order to restart a computation of~$\mach{A}$),
and then recover the position~$\length{u}$.
To overcome this issue,
we use annotations,
so that it is possible to simulate computations of~$\mach{A}$ without moving the head more than~$2n$ cells to the left of position~$\length{u}$.
In this way, the automaton can recover the position~$\length{u}$ by maintaining,
in its finite control,
the distance of the head from that position.

Let~$w\in\ialphab*$.
The idea is to consider an annotated word~$x\in{(\ialphab\times\set{0,1})}^*$
such that ${\pi_1(x)=w}$
and the following property holds. 
Logically dividing~$x$ into factors of length~$n$ (or possibly less for the last one),
the~$i$\=/th factor encodes~$\QX[A]{w[0,i\cdot n-1]}$.
Formally,
decomposing~$x$ as~$x=x_1\cdots x_{k+1}$ with~$\length{x_{k+1}}\leq n$ and~$\length{x_i}=n$ for each~$i\leq k$,
every~$\pi_2(x_i)$ (except, possibly~$\pi_2(x_{k+1})$ if it has length less than~$n$)
encodes the set~$\QX[A]{\pi_1(x_1\cdots x_i)}=\QX[A]{w[0,i\cdot n-1]}$.
In this way,
our nondeterministic procedure~\nsimulA
starting from some position~$i\cdot n+j$, ${0\leq i\leq k}$ and~$0\leq j<n$,
operates in two phases.
First, it scans~$\pi_2(x_i)$,
from which it extracts some nondeterministically-chosen state~$\p\in\QX[A]{w[0,i\cdot n-1]}$
--- for~$i=0$, we assume~$x_i=\lend$ from which the machine extracts~$\p=\qstart$.
Second, it performs a direct simulation of~$\mach{A}$ on~$\pi_1(x_{i+1}[0,j])$,
starting from the selected state~$\p$
and halting as soon as the cell at position~$i\cdot n+j$ is entered.
At that time, the reached state in the simulated path,
belongs to~$\delta(\p,w[i\cdot n,i\cdot n+j-1])$,
and thus to~$\QX[A]{w[0,i\cdot n+j-1]}$ since~${\p\in\QX[A]{w[0,i\cdot n-1]}=\delta(\qstart,w[0,i\cdot n-1])}$.
\smallbreak

A subset~$S$ of~$\Q=\interv{n}$ is naturally encoded as a length\=/$n$ word~$\defem{\encode{S}\in\set{0,1}*}$ as follows:%
\begin{equation}
	\encode{S}[\p] = 1
	\qquad\iff\qquad
	\p\in S
	\text.
\end{equation}
Hence,
provided~${\pi_2(x_i)=\encode{\QX[A]{w[0,i\cdot n]}}}$,
in order for~\nsimulA to select~${\p\in\QX[A]{w[0,i\cdot n]}}$
during its first phase,
it is sufficient to choose a position~$\p$ of~$x_i$ such that~$\pi_2(x_i[\p])=1$.
The annotation of a word~$w\in\ialphab*$ is defined now.
\begin{definition}
	\label{def:1nfa/annot}
	For each~$w\in\ialphab*$,
	we let~\defem{$\mathbf{a}_w$} be the $\length{w}$\=/length word over~$\set{0,1}$ defined as follows.
	Let~$k$ and~$r$ be such that~$\length{w}=kn+r$ with~$0\leq r<n$:
	\begin{itemize}[nosep]
		\item if~$k=r=0$ (\ie, $w=\ew$) then~$\mathbf{a}_w=\ew$, otherwise,
		\item if~$r=0$ then~$\mathbf{a}_w=zy$,
			where~$z=\mathbf{a}_{w[0,(k-1)n]}$,
			and~$y=\encode{\QX[A]{w}}$,
		\item if~$r>0$ then~$\mathbf{a}_w=zy$,
			where~$z=\mathbf{a}_{w[0,kn]}$,
			and~$y=0^{r}$.
	\end{itemize}
	The word~$\annot{w}$ is the word~$x$ over~$\analphab=\ialphab\times\set{0,1}$
	such that~$\pi_1(x)=w$ and~$\pi_2(x)=\mathbf{a}_w$.
\end{definition}

Our goal is to design a \twnfa~$\mach{B}$ over~$\analphab=\ialphab\times\set{0,1}$
of polynomial size in~$n$,
with two distinguished states~$\qacc$ (for acceptance) and~$\qrej$ (for rejection)
that satisfies:
\begin{itemize}
	\item $\mach{B}$ accepts~$x$ \ifof $\pi_1(x)\in\langof{A}$ and~$x=\annot{\pi_1(x)}$;
	\item $\mach{B}$ rejects~$x$ \ifof $\pi_1(x)\notin\langof{A}$ and~$x=\annot{\pi_1(x)}$.
\end{itemize}
\begin{equation}
\label{selfverifying}
\tag{$D$}
\begin{gathered}
	\text{$\mach{B}$ accepts~$x$ \ifof $\pi_1(x)\in\langof{\mach{A}}$ and~$x=\annot{\pi_1(x)}$;}
	\\
	\text{$\mach{B}$ rejects~$x$ \ifof $\pi_1(x)\notin\langof{\mach{A}}$ and~$x=\annot{\pi_1(x)}$.}
\end{gathered}
\end{equation}
Notice that~$\mach{B}$ itself is not self-verifying,
as it can neither accept nor reject an input~$x$ such that~$x\neq\annot{\pi_1(x)}$.
However, the \twnfacg~$\struct{\mach{B},\ialphab,\set{0,1}}$
is self-verifying since every input~$w\in\ialphab*$ admits an annotated variant~$\annot{w}$
that should be either accepted or rejected by~$\mach{B}$ according to whether~$w$ belongs to~$\langof{A}$ or not.

The design of~$\mach{B}$ follows the above-explained inductive counting strategy.
In particular, it uses the already-presented procedures \enumQX, \memberQX, and~\countnextQX
(see \cref{proc:enumQX,proc:memberQX,proc:countnextQX}),
as well as~\nsimulA.
To implement the latter,
$\mach{B}$ maintains two variables in its finite control:
the value of its head position modulo~$n$
(so it knows where each factor~$x_i$ begins, and to which state~$\p$ an annotation symbol~$1$ correspond),
and the distance of its current head position to the position~$\length{z}$,
where~$z$ is the input prefix under consideration.
By the locality of~\nsimulA,
the latter information is an integer less than~$2n$.

Not only~$\mach{B}$ has to behave correctly on inputs~$\annot{w}$:
it also has to detect ill-formed inputs,
namely words~$x\in\analphab*$ for which~$x\neq\annot{\pi_1(x)}$.
To this end, each time the length of the prefix~$z$ under consideration is a positive multiple of~$n$,
before considering the next prefix,
$\mach{B}$ checks the annotation of the length\=/$n$ suffix~$x_i$ ($i={\length{z}}/{n}$) of~$z$.
This can easily be done,
since at that time~$m=\card{\QX[A]{\pi_1(z)}}$ has been computed.
Hence, the automaton can check that~$\pi_2(x_i)$ has~$m$ occurrences of~$1$,
and, using~\enumQX,
that~$\pi_2(x_i[\p])=1$ for each~$\p\in\QX[A]{\pi_1(z)}$.

Evaluating the size of~$\mach{B}$,
we get that~$\bigO{n^3}$ states are sufficient
for implementing \nsimulA,
including the two above-mentioned state components relative to the head position.
Next, $\mach{B}$ further uses a $\bigO{n^4}$\=/state component
for storing~$\var{m}=\card{\QX[A]{u}}$, $\var{m_{next}}\leq\card{\QX[A]{u\tau}}$, and two intermediate states~$\var{\q}$ and~$\var{\q_{prev}}$.
Hence, the total number of states belongs to~$\bigO{n^7}$.
\begin{theorem}
	\label{thm:1nfa to sv1-la}
	Every $n$\-/state \ownfa
	has an equivalent self-verifying \twnfacg
	with~$\bigO{n^7}$ many states and~$2$ annotation symbols.
\end{theorem}

A direct consequence is that complementing \las
costs at most a single exponential.
\begin{corollary}
	\label{cor:complement 1-la}
	For each $n$\=/state \la recognizing some language~$L\subseteq\ialphab*$,
	there exists a \la with a single exponential number of states in~$n$
	and~$3\card{\ialphab}$ work symbols
	which recognizes the complement of~$L$.
\end{corollary}
\begin{proof}
	From \cite[Theorem 2]{PP14}, each $n$-state \la over~$\ialphab$
	can be simulated by a \ownfa with at most~${n2^{n^2}}$ states.
	Also by \cref{thm:1nfa to sv1-la}, each $m$-state  \ownfa can be converted into
	an equivalent self-verifying \twnfacg (which is a particular case of \la) with $\bigO{m^7}$ states
	and~$2$ annotation symbols.
	Combining these two results, one can simulate each $n$-state \la,
	by a self-verifying \la with work alphabet~$\walphab=\ialphab\cup(\ialphab\times\set{0,1})$
	and a number of states in~${\bigO{2^{7(n^2+\log{n})}}\subset 2^{\bigO{n^2}}}$.%
\end{proof}
As demonstrated in~\cite{GPT25},
the exponential bound in the above corollary cannot be avoided in the worst case,
even when the source device is a \emph{unary \twdfacg},
namely a \twnfacg over a singleton input alphabet
whose underlying \twnfa (working on nonunary annotated words) is \emph{deterministic}.%
\footnote{Remark that~\twdfacgs are nondeterministic devices, since the annotation phase is nondeterministic.}
\end{document}

\section{Complementing \twnfas with \twnfacgs}
\label{sec:2nfa}

In this section,
we show how to simulate an arbitrary $n$\=/state \twnfa~$\mach{A}$ over~$\ialphab$
with a self-verifying \twnfacg~$\mach{M}$
that has polynomially many states in~$n$
and annotation alphabet~$\set{0,1}$.
In order for~$\mach{M}=\struct{\mach{B},\ialphab,\set{0,1}}$ to be self-verifying,
%
we shall ensure that its underlying \twnfa~$\mach{B}$ over~$\analphab=\ialphab\times\set{0,1}$
satisfies the following sufficient property
(see \cref{res:2nfa checking annotation and testing membership}).
For every word~$w\in\ialphab*$,
there exists a word~${\annot{w}\in\analphab*}$
such that~${\pi_1(\annot{w})=w}$, and on input~$x\in\analphab*$:
\begin{itemize}
	\item $\mach{B}$ \emph{accepts}~$x$ \ifof $\pi_1(x)\in\langof{A}$ and~$x=\annot{\pi_1(x)}$;
	\item $\mach{B}$ \emph{rejects}~$x$ \ifof $\pi_1(x)\notin\langof{A}$ and~$x=\annot{\pi_1(x)}$.
\end{itemize}
Notice that~$\mach{B}$ itself is not a self-verifying \twnfa.
Indeed, on inputs not belonging to~$\annot{\ialphab*}$,
$\mach{B}$ can neither accept nor reject.%
\footnote{%
	$\mach{B}$ is actually a \emph{don't-care} automaton~\cite{MPR17},
	namely an automaton with the self-verifying property
	on inputs from a restricted domain (here, inputs of the form~$\annot{w}$ for some~$w\in\ialphab*$).%
}
Yet, as the goal is to check whether a word~$w\in\ialphab*$ belongs to~$\langof{A}$,
and since each such word has an annotated version~$\annot{w}$ that is either accepted or rejected by~$\mach{B}$,
we obtain self-verifyingness for the \NOYES{\la}{\twnfacg}~$\mach{M}$.

The section is structured as follows.
In \cref{sec:2nfa/ltables},
we recall the principle of Shepherdson's construction~\cite{She59},
introducing the basic concepts and properties on which such a simulation as much as ours rely.
The definition of~$\annot{w}$ is given at the end of the section.
The automaton~$\mach{B}$ is then designed in \cref{sec:2nfa/B}
and the main result is stated.

\subsection{\Ltables}
\label{sec:2nfa/ltables}
In~\cite{She59}, Shepherdson proposed a construction to simulate \twdfas by \owdfas,
which has then been generalized to the simulation of \twnfas and even \las by \owdfas,
see, \eg, \cite{Kap05,PP14}.
The main ingredient of this construction
is to store in each state of the finite control of the simulating \owdfa,
a table, that we call~\emph{\Ltable},%
\footnote{L for ``Left''; In the literature, they have sometimes been called ``\emph{Shepherdson's tables}''.}
describing the finitely-many possible behaviors of the simulated two-way machine
that may occur on the portion of the tape to the left of the current head position. 
This is formalized in the following.

For~$u\in\ialphab*$,
an \defem{\Lseg over~$u$}[\wrt~$\mach{A}$]
is a computational path of~$\mach{A}$ over~$\lend uv\rend$ for some~$v\in\ialphab*$
that starts from tape position~$\length{u}$ (hence reading the last symbol of~$\lend u$)
and ends in position~$\length{u}+1$ (hence, reading the first symbol of~$v\rend$)
visiting only positions~$j\leq\length{u}$ in the meantime (hence, independent from~$v$).
The \defem{\Ltable of~$u$}[\wrt~$\mach{A}$],
denoted~\defem{\lt[\mach{A}]{u}},
is the set of pairs~$(\p,\q)$ such that
there exists an \Lseg over~$u$ starting in state~$p$ and ending in state~$q$.
Formally:
\begin{align*}
	\lt[A]{u}&=\set{(\p,\q)\in\Q^2}[\cfg{x}{\p}{\sigma}\yield*[A]\cfg{\lend u}{\q}{}]\text,
	&
	\text{where~$x\sigma=\lend u$ with~$\length{\sigma}=1$}
	\text.
\end{align*}
Observe that there are finitely many\xspace%
such tables.
%
Also, for every~$u\in\ialphab*$,
we denote by~\defem{${\QX[A]{u}}$} the set of states that~$\mach{A}$
may enter when it visits the cell containing the right endmarker for the first time
during a computation over input~$u$.
Namely,
\begin{align*}
	\QX[A]{u}&=\set{\q\in\Q}[\cfg{\lend}{\qstart}{u}\yield*[A]\cfg{\lend u}{\q}{}]
	\text.
\end{align*}
Again, remark that there are finitely many such sets.
Shepherdson observed that
knowing~$\QX[A]{u}$ and~$\lt[A]{u}$ (but not~$u$)
is sufficient for deciding whether~$u\in\langof{A}$.
In order to simplify,
we slightly modify the simulated \twnfa,
so that knowing~$\lt[A]{u}$ will be sufficient to recover~$\QX[A]{u}$
and thus the acceptance of~$u$.
\begin{lemma}
	\label{res:ltables only}
	Each $n$\=/state \twnfa~$\mach{A}$ over~$\ialphab$
	admits an equivalent ${(n+1)}$\=/state \twnfa~$\mach{A'}$
	with an inaccessible distinguished state~$\qrestart$
	such that,
	for each~$u\in\ialphab*$ and each state~$\p$ of~$\mach{A'}$,
	on the one hand~$(\qrestart,\p)\in\lt[A']{u}$ \ifof~$\p\in\QX[A']{u}$,
	and on the other hand~$(\p,\qrestart)\notin\lt[A']{u}$.
	Furthermore,
	$\delta'(\qrestart,\rend)=\set{(\qrestart,\lmove)}$ where~$\delta'$ is the transition function of~$\mach{A'}$.
\end{lemma}
\begin{proof}
	The \twnfa~$\mach{A'}$ is obtained from~$\mach{A}$ by
	\begin{enumerate*}[label={(\arabic*)}, itemjoin={,\xspace}, itemjoin*={, and\xspace}]
		\item adding a new state~$\qrestart$
		\item adding transitions returning~$(\qrestart,\lmove)$ from~$\qrestart$ on every symbol~$\sigma\in\ialphab\cup\set{\rend}$
		\item adding a transition from~$\qrestart$ returning~$(\qstart,\rmove)$ on~$\lend$
			where~$\qstart$ is the initial state of both~$\mach{A'}$ and~$\mach{A}$.%
	\end{enumerate*}%
	\footnote{Remarkably, although useless for our purpose, the transformation preserves determinism.}
	Hence, from all configurations of the form~$\cfg{x}{\qrestart}{y}$ with~$xy=\lend w\rend$,
	the machine deterministically reaches the configuration~${\cfg{\lend}{\qstart}{w\rend}}$,
	namely the initial configuration of both~$\mach{A'}$ and~$\mach{A}$,
	from where it performs a direct simulation of~$\mach{A}$.
	In particular,~$\lt[A']{u}(\qrestart)=\QX[A']{u}=\QX[A]{u}$ for all~$u\in\ialphab*$.
	Also, since the only transitions entering~$\qrestart$ are left-move looping around~$\qrestart$,
	$\qrestart$ is inaccessible, and~$(\p,\qrestart)\notin\lt[A']{u}$ for every~$u$ and~$\p$.
\end{proof}

The key of Shepherdson's construction
is that, given~$\lt[A]{u}$ and~$\tau\in\ialphab$,
it is possible to compute~$\lt[A]{u\tau}$ without accessing~$u$.
Indeed, an \Lseg on~$u\tau$ can be decomposed as a sequence
alternating left-moves over~$\tau$ and~\Lsegs on~$u$,
followed by a final right-move over~$\tau$.
This is formalized in \cref{res:update ltables} below,
using the following notions.
For~$\tau\in\ialphab\cup\set{\rend}$ and~$k\geq0$,
we define the binary relation~\defem{$\lT[A]{u\tau}{k}$} (\resp~\defem{$\lS[A]{u\tau}{k}$}) on~$\Q$
as the set of pairs~$(\p,\q)$ for which
there is a computational path
that starts at position~$\length{u}+1$ in state~$\p$,
ends at the same position in state~$\q$,
visits only positions~$j\leq\length{u}+1$ in the meantime,
and visits the position~$\length{u}+1$ exactly (\resp at most)~$k+1$ times
(including the initial and final visits, in state~$\p$ and~$\q$, respectively).
\begin{definition}
	\label{def:T and S}
	\label{def:T}
	\label{def:S}
	Let~$\mach{A}$ be a \twnfa over~$\ialphab$,
	$u\in\ialphab*$, $\tau\in\ialphab\cup\set{\rend}$, and~$k\geq0$.
	We define:
	\begin{align*}
		\lT[A]{u\tau}{0}&=\set*{(\p,\p)}[\p\in\Q%
		\vphantom{\exists\q[s],\r\in\Q\mit*{such that}(\p,\q[s])\in\lT[A]{u\tau}{k},\;(\r,\lmove)\in\delta(\q[s],\tau)\mit*{and}(\r,\q)\in\lt[A]{u}}]
		\\
		\lT[A]{u\tau}{k+1}&= \set*{(\p,\q)}[\exists\q[s],\r\in\Q\mit*{such that}(\p,\q[s])\in\lT[A]{u\tau}{k},\;(\r,\lmove)\in\delta(\q[s],\tau)\mit*{and}(\r,\q)\in\lt[A]{u}]
		\text.
	\end{align*}
	Furthermore we define~$\lS[A]{u\tau}{k}=\bigcup_{j=0}^{k}\lT[A]{u\tau}{j}$
	and~$\lS[A]{u\tau}{*}=\bigcup_{j\in\Nat}\lT[A]{u\tau}{j}$.
\end{definition}
\begin{remark}
	\label{rk:T1 from t}
	By definition,~$\lS[A]{u\tau}*$ is the reflexive and transitive closure of~$\lT[A]{u\tau}{1}$
	which is characterized by:
	\begin{align*}
		\label{eq:T1 from t}
		(\p,\q)&\in\lT[A]{u\tau}{1}
		&&\iff&&
		\exists\r\in\Q\mathinner{:}\;
		(\r,\q)\in\lt[A]{u}
		\mit{and}
		(\r,\lmove)\in\delta(\p,\tau)
		\text.
	\end{align*}
\end{remark}
Since the family~${(\lS[A]{u\tau}{k})}_{k}$ is an increasing sequence of subsets of~$\Q^2$ \wrt inclusion,
it is ultimately constant.
\begin{proposition}
	\label{res:iterated ltables bound}
	Let~$\mach{A}$ be an~$n$\=/state \twnfa over~$\ialphab$,
	$u\in\ialphab*$, and~$\tau\in\ialphab\cup\set{\rend}$.
	Then \({
			\lS[A]{u\tau}{*}=\lS[A]{u\tau}{j}
		}\)
	for all~$j\geq n(n-1)$.
\end{proposition}
\begin{proof}
	By definition, for all~$j\geq 0$,
	\(
		\lS[A]{u\tau}{j}\subseteq\lS[A]{u\tau}{j+1}\subseteq\lS[A]{u\tau}*\subseteq\Q^2
		\text.
	\)
	Furthermore, by the inductive definition of~$\lT[A]{u}{k}$,
	if~$\lS[A]{u}{k}=\lS[A]{u}{k+1}$ for some~$k$
	then~$\lS[A]{u}{k}=\lS[A]{u\tau}*$.
	Since~$\card{\lS[A]{u\tau}{0}}=n$ and~$\card{\lS[A]{u\tau}{*}}\leq n^2$,\xspace%
	\footnote{%
		If~$\mach{A}$ is in the form of \cref{res:ltables only},
		then\xspace%
		$\card{\lS[A]{u\tau}*}\leq n(n-1)+1$ and thus~$\lS[A]{u\tau}*=\lS[A]{u\tau}{n(n-2)+1}$.%
	}
	this surely happens for~$k\leq n(n-1)$, and hence~$\lS[A]{u\tau}*=\lS[A]{u\tau}{j}$ for all~$j\geq n(n-1)$.%
\end{proof}
In order to build~$\lt[A]{u\tau}$ from~$\lt[A]{u}$,
we rely on the following property.
\begin{proposition}
	\label{res:update ltables}
	\label{res:t from S*}
	Let~$\mach{A}={\struct{Q,\ialphab,\delta,\qstart,\qfin}}$ be a \twnfa,
	$u\in\ialphab*$, and~$\tau\in\ialphab$.
	Then ${(\p,\q)\in\lt[A]{u\tau}}$ \ifof
	there exists~$\r$
	such that~${(\p,\r)\in\lS[A]{u\tau}*}$ and~$(\q,\rmove)\in\delta(\r,\tau)$.
\end{proposition}
A direct consequence of the above proposition is that~$\lt[A]{u}=\lt[A]{v}$ implies~$\lt[A]{uw}=\lt[A]{vw}$ for every~$w$.
Also, provided~$\mach{A}$ is in the form of \cref{res:ltables only},
we can decide whether~$w\in\langof{A}$ with the only information of~$\lS[A]{w\rend}{*}$,
which is determined from~$\lt[A]{w}$:
\begin{proposition}
	\label{res:ltables for recognition}
	Let~$\mach{A}$ be a \twnfa over~$\ialphab$ in the form of \cref{res:ltables only},
	and~$w\in\ialphab*$.
	Then~$w\in\langof{A}$ \ifof
	$(\qrestart,\qfin)\in\lS[A]{w\rend}*$
	where~$\qfin$ is the accepting state of~$\mach{A}$.
\end{proposition}
\begin{proof}
	By definition,
	a word~$w\in\ialphab*$ belongs to~$\langof{A}$
	\ifof there exists a computational path
	that starts from the initial configuration
	and halts in the accepting state~\qfin with the head scanning the right endmarker.
	Every such path should thus visit the right endmarker at least once.
	Hence, $w\in\langof{A}$ \ifof there exists~$\p$
	such that~$\p\in\QX[A]{w}$ and $(\p,\qfin)\in\lS[A]{w\rend}*$.
	Since~$\QX[A]{w}=\lt[A]{w}(\qrestart)$ and~$\delta(\qrestart,\rend)=\set{(\qrestart,\lmove)}$,
	this is equivalent to~$(\qrestart,\qfin)\in\lS[A]{w\rend}*$.%
\end{proof}

Not every binary relation~$R\subseteq\Q^2$
is equal to~$\lt[A]{u}$ for some~$u$.
Nevertheless, each~$R\subseteq\Q^2$ can be updated according to \cref{res:update ltables}.
We formalize this fact in the following,
by introducing a variant~$\mach{A}_{/R}$ of~$\mach{A}$ so that~$\lt[A_{/R}]{\ew}=R$.
\begin{definition}
	\label{def:A/R}
	Let~$\mach{A}$ be a \twnfa with state set~$\Q$,
	and let~$R\subseteq\Q^2$.
	We define~\defem{$\mach{A}_{/R}$}
	as the \twnfa obtained from~$\mach{A}$
	by overwriting its transitions on~$\lend$ according to~$R$.
	More precisely, $\mach{A}_{/R}$ has the same transitions as~$\mach{A}$ on symbols distinct from~$\lend$,
	and transitions from~$\p$ to~$(\q,\rmove)$ on~$\lend$ \ifof~$(\p,\q)\in R$.
\end{definition}
Trivially, $\lt[\mach{A}_{/R}]*{\ew}=R$ and~$\mach{A}_{/\lt[A]{\ew}}=\mach{A}$.
Also, if~$R=\lt[A]{u}$ for some~$u$ then~$\lt[A_{/R}]{v}=\lt[A]{uv}$ for every~$v$.
\medbreak

In Shepherdson's construction, after reading a prefix~$u$ of the input,
the simulating \owdfa stores the whole table~$\lt[A]{u}$
(along with the set~$\QX[A]{u}$,
which thanks to \cref{res:ltables only} is not needed in our presentation)
in its finite control.
Then, on symbol~$\tau$, it updates it to~$\lt[A]{u\tau}$ according to~\cref{res:update ltables}.
Finally, it decides acceptance according to~\cref{res:ltables for recognition}.
This method comes with a cost:
storing the \Ltables in the finite control
implies an exponential number of states
(which, for the simulation of \twnfas by \owdfas, cannot be avoided in general; see~\cite{Kap05} for a precise analysis).
In order to keep the size of our simulating self-verifying \twnfacg polynomial,
we do not store the successive \Ltables in the finite control of the machine.
We rather maintain their sizes as a state component,
and encode \emph{some} of them \emph{on the tape}.
Intuitively, the tape will be virtually divided into portions of length~$n^2$
(possibly the last portion being shorter),
so that the~$j$\=/th portion stores on its annotation track
an encoding of the table~$\lt[A]{x}$ where~$x$ is the input prefix of length~$jn^2$.
We naturally encode a relation~$R\subseteq\Q^2$ in a length\=/$n^2$ word~\defem{$\encode{R}$} over~$\set{0,1}$, as follows:%
\footnote{%
	Since the only relations over~$\Q$ we encode are relations~$\lt[A]{u}$ for some~$u$,
	and because they are included in~$\Q\times(\Q\setminus\set{\qrestart})$
	when assuming~$\mach{A}$ normalized according to~\cref{res:ltables only},
	a word of length~$n(n-1)$ would be enough for encoding it.
	This would match the space needed for storing the pairs~$(\lt[A]{u},\QX[A]{u})$
	(of the corresponding non-normalized $(n-1)$\=/state \twnfa)
	in Shepherdson's construction.
	For readability, we do not apply this minor optimization.%
}
\begin{align*}
	\encode{R}[\p n+\q]=1
	&\:\iff\:
	(\p,\q)\in R
	&\text{%
		for each~$\p,\q\in\Q=\interv{n}$%
	}
	\text.
\end{align*}
Notice that the encoding defines a bijection between binary relations on~$\Q$
and length\=/$n^2$ words over~$\set{0,1}$.
We denote by \defem{\decodename} the inverse of~\encodename,
\ie, $\decodename=\encodename^{-1}$.
\begin{definition}
	\label{def:annot}
	For~$w\in\ialphab*$,
	\defem{the annotation of~$w$}
	is the word~${\defem{\mathsf{a}_{w}}\in\set{0,1}*}$
	of length~$\length{w}$
	defined as follows.
	Let~$k\in\Nat$ and~$r\in\interv{n^2}$ such that~${\length{w}=kn^2+r}$.
	\begin{itemize}[nosep]
		\item If~$k=r=0$ (\ie,~$w=\ew$) then  ${\mathsf{a}_{w}=\ew}$, otherwise,
		\item If~$r=0$ then ${\mathsf{a}_{w}=zy}$ where~${z=\mathsf{a}_{w[0,(k-1)n^2]}}$ and~${y=\encode{\lt[A]{w}}}$,
		\item If~$r>0$ then ${\mathsf{a}_{w}=zy}$ where~${z=\mathsf{a}_{w[0,kn^2]}}$ and~${y=0^r}$.
	\end{itemize}
	The word~$\annot{w}$ is the word~$x$ over~$\analphab=\ialphab\times\set{0,1}$
	such that~$\pi_1(x)=w$ and~$\pi_2(x)=\mathbf{a}_w$.
\end{definition}

\subsection{The automaton~$\mach{B}$}
\label{sec:2nfa/B}
As in \cref{sec:1nfa},
a key ingredient in our construction
is a subprocedure, implemented by a \twnfa, which,
starting from and ending in some position~$\length{v}$ for some prefix~$v$ of the input,
nondeterministically simulates the computational paths of~$\mach{A}$ on~$\lend v$.
Unlike those of \ownfas described in \cref{sec:1nfa},
such computational paths are here \Lsegs or variants of \Lsegs.
Again, this procedure is made possible by the use of annotations,
which allow to keep the head close to the position~$\length{v}$
during the simulation.
This idea has already been used in~\cite{GP19}\xspace
for proving a polynomial upper bound
for the simulation of~\las by halting~\twnfacgs.
However, in that construction the resulting machine is only able to check
the inclusion of the encoded relations in the corresponding \Ltables.
In other words, the simulating \twnfacg is allowed to \emph{``lose''} some pairs of the \Ltables.
Although this is sufficient for the simulated machine to recover acceptance of the input,
it turns out to be insufficient if, as in the present work, we aim to recover \emph{rejection} of the input.
To address this lack of information,
following the same strategy as in \cref{sec:1nfa} and~\cite{GMP07,GGP14},
our construction uses inductive counting to check that
every encoded relation is \emph{equal} to the corresponding \Ltable,
and finally detect whether the input belongs to~$\langof{A}$ or not.

\subsubsection{Simulating \Lsegs using annotations.}
\label{sec:nsimul}
\label{sec:nmember}
\label{sec:sliding window}
The automaton~$\mach{B}$
maintains a variable~$\hp$ ranging over~$\interv{2n^2}$ in its finite control,
which is updated according to each head move as now explained.
The variable~$\hp$ is incremented on right moves and decremented on left moves
like a counter
with the two following differences:
decrementing from value~$0$ is forbidden (hence left-moves from a state in which~$\hp=0$ are forbidden),
and incrementing from value~$2n^2-1$ resets the counter to~$n^2$.
In the initial configuration, in which the head is at position~$1$, $\hp=n^2$.
Hence, at any point of the computation with the head at position~$h$,
the value of~$\hp$ is congruent to~$h-1$ modulo~$n^2$.
We call \defem{current window} the portion of the tape of length at most~$2n^2$,
going from position~$\max(0,h-\hp)$ to position~$\min(\ell+1,h-\hp+2n^2-1)$ where~$\ell$ is the input length.
By \defem{relative position~$i$}, for~$i\in\interv{2n^2}$,
we refer to the position~$i$ relatively to the current window,
\ie, the absolute position~$\max(0,h-\hp)+i$.
The window will slide from left to right along the input.
Indeed, on the one hand,
as decrementing~$\hp$ from value~$0$ is forbidden
the head can never visit cells to the left of the current window.
On the other hand,
updating the value of~$\hp$ from~$2n^2-1$ to~$n^2$ on a right-move
shifts the window to the right by~$n^2$ cells
in the following sense:
after the shift,
a cell that was at some relative position~$i$ before the shift
is either not covered by the window if~$i<n^2$, or at relative position~$i-n^2$ otherwise.
By using~$\hp$, $\mach{B}$ can navigate within this window without getting lost.
From now on,
we assume the machine has always access to the value~$\hp$
without mentioning it explicitly.
More generally, we say that a \twnfa is \defem{$\hp$\=/aware} when it has access to~\hp,
and we do not count the underlying state component when analyzing its size.
In a (partial) configuration of some $\hp$\=/aware \twnfa,
\defem{$\hpQ{\q}{i}$} indicates state~$\q$ with~$\hp=i$\xspace%
.

One of the key ingredients allowing our inductive counting
is a nondeterministic subroutine \nsimult (\cref{proc:nsimult}),
which,
assuming the annotation on the left side of the current window is correct,
allows to simulate \Lsegs of~$\mach{A}$
while keeping the head within the current window.
In particular, on input~$w$, if called from position~$\length{x}$ for a prefix~$x$ of~$\annot{w}$,
the procedure eventually halts (if an \Lseg on~$\pi_1(x)$ is found) with the head at position~$\length{x}+1$.
The procedure can be implemented by a \twnfa of polynomial size in~$n$, as stated in the following \lcnamecref{res:nsimult}
(recall \cref{def:A/R}).
\begin{lemma}
	\label{res:nsimult}
	Let~$i\in\interv{n^2}$.
	There exists a \hp\=/aware \twnfa~$\mach{C}_i$ with same state set~$\Q$ as~$\mach{A}$,
	such that,
	for every~${x\in\set{\lend}\cup\analphab^{n^2}}$,
	every~$y\in\analphab^i$,
	and every~$\p,\q\in\Q$:%
	\begin{align*}
		\cfg{z}{\hpQ*{\p}{n^2+i-1}}{\sigma}	&\yield*[C_i]	\cfg{z\sigma}{\hpQ*{\q}{n^2+i}}{}
		&\iff&&
		(\p,\q)	&\in	\lt[\mach{A}_{/R}]*{\pi_1(y\sigma)}
	\end{align*}
	where~$z\sigma=xy$ with~$\length{\sigma}=1$,
	and~$R$ equals~$\lt[A]{\ew}$ if~$x=\lend$ or~$\decode{\pi_2(x)}$ otherwise.
	Furthermore, $\mach{C}_i$ has no transition outgoing configurations in which~$\hp\geq n^2+i$;
	in particular, the above configuration ${\cfg{z\sigma}{\hpQ{\q}{n^2+i}}{}}$ is halting.
\end{lemma}
\begin{procedure}[tb]
	\caption{%
		()\space%
		\nsimult{$\p$}
		\label{proc:nsimult}%
	}
	%
	$\var{i}\gets\hp-n^2+1$
	\label{l:nsimult/i}\;
	$\var{\p_{curr}}\gets p$\;
	\While(\label{l:nsimult/main loop cond}){$\hp<n^2+i$}{%
		\eIf(\label{l:nsimult/direct simul cond}){%
			$\hp\geq n^2$
			\Or $\read{}=\lend$%
		}%
		{%
			\tcp{direct simulation of~$\mach{A}$ reading the input track}
			choose~$(\r,d)\in\delta(\var{\p_{curr}}, \pi_1(\read{}))$
			\label{l:nsimult/direct simul begin}\;
			$\var{\p_{curr}}\gets\r$\;
			move the head according to~$d$
			\label{l:nsimult/direct simul end}\;
		}
		{%
			\tcp{recovering \Lseg from the annotation track}
			\While(\label{l:nsimult/scan table begin}){\True}{%
				\If(\label{l:nsimult/scan table interval}){$\var{\p_{curr}}\cdot n\leq\hp<(\var{\p_{curr}}+1)n$ \And~$\pi_2(\read{})=1$}{%
					choose~$b\in\set{\True,\False}$
					\label{l:nsimult/scan table choice}\;
					\lIf{$b$}{\Break}
				}
				\lIf(\label{l:nsimult/scan table abort}){$\hp=0$}{\Abort}
				move the head leftward\;
			}
			$\var{\p_{curr}}\gets\hp\mod n$
			\label{l:nsimult/scan table update}\;
			move the head rightward to relative position~$n^2$
			\label{l:nsimult/scan table move}
			\label{l:nsimult/scan table end}\;
		}
	}
	\Return $\var{\p_{curr}}$\;
\end{procedure}
\begin{proof}
	The behavior of~$\mach{C}_i$ is described in~\cref{proc:nsimult},
	where the value of~$i$ is deduced from the initial relative position (\cref{l:nsimult/i}),
	and which uses a variable~$\var{\p_{curr}}$ ranging over~$\Q$.
	The states of~$\mach{C}_i$ are the valuations of this variable.
	The absence of outgoing transitions when~${\hp\geq n^2+i}$ is ensured by the main loop condition (\cref{l:nsimult/main loop cond}).
	Let~$x$, $y$, $z$, $\sigma$, and~$R$ be as in the lemma statement.
	We also let~$z'\sigma'=\lend\pi_1(y)$ with~$\length{\sigma'}=1$.
	Starting from the last position of~$xy=z\sigma$ (thus scanning~$\sigma$) with~$\var{\p_{curr}}=\p$ and~$\hp=n^2+i-1$,
	$\mach{C}_i$ nondeterministically simulates a computational path of~$\mach{A}_{/R}$ starting from configuration~${\cfg{z'}{\p}{\sigma'}}$.
	Yet, $\mach{C}_i$ does not work on~$z'\sigma'$ but on~$xy$.\xspace
	Hence, in order to simulate~$\mach{A}_{/R}$,
	$\mach{C}_i$ proceeds in two modes
	(mainly distinguished by whether~$\hp\geq n^2$ or not, \cf \cref{l:nsimult/direct simul cond}).
	\begin{enumerate}[leftmargin=0pt,label={Mode~$\arabic*$.},font=\strongsty,itemindent=\widthof{\strong{Mode~$2$.\space}}+\parindent]
		\item\label{nsimult:mode/direct simul}
			As far as it scans the portion containing~$y$ ($\hp\geq n^2$),
			it performs a direct simulation by reading the symbols from the input track.
			Since on this portion these symbols are distinct from~$\lend$,
			the transition of~$\mach{A}_{/R}$ are the same as those of~$\mach{A}$.
			Hence, $\mach{C}_i$ nondeterministically chooses one transition of~$\mach{A}$ on the corresponding symbol,
			and then updates its state and moves its head accordingly
			(\crefrange{l:nsimult/direct simul begin}{l:nsimult/direct simul end}).
		\item\label{nsimult:mode/read table}
			As soon as the last position of the portion containing~$x$ is entered
			(from the right, detected as~$\hp<n^2$),
			it simulates a transition of~$\mach{A}_{/R}$ on~$\lend$.
			If~$x=\lend$ then~$R=\lt[A]{\ew}$ and thus such transitions are the same as those of~$\mach{A}$.
			Hence the same simulation as in the previous case works
			(\cf the second condition for entering the \hyperref[nsimult:mode/direct simul]{previous mode} on \cref{l:nsimult/direct simul cond}).
			Otherwise, $\mach{C}_i$ scans backward the annotation track carrying~$\pi_2(x)$
			(\crefrange{l:nsimult/scan table begin}{l:nsimult/scan table end})
			in order to find some relative position~$\p n+\q$
			where~$\p$ is the current value of~$\var{\p_{curr}}$
			(detected as~$\var{\p_{curr}}\cdot n\leq\hp<(\var{\p_{curr}}+1)n$
			and such that~$\pi_2(x[\p n+\q])=1$ (\cref{l:nsimult/scan table interval}).
			Such a position indeed indicates that~$(\p,\q)$ belongs to~$R$,
			or equivalently,
			that from state~$\p$ scanning~$\lend$, $\mach{A}_{/R}$ may move its head rightward and enter~$\q$.
			Upon encountering such a position,
			$\mach{C}_i$ nondeterministically chooses either to ignore it or to select it
			(\cref{l:nsimult/scan table choice}).
			In the former case it proceeds with the backward scan of~$\pi_2(x)$,
			while in the latter it sets~$\var{\p_{curr}}$ to~$\q$
			and moves the head rightward to relative position~$n^2$
			(\crefrange{l:nsimult/scan table update}{l:nsimult/scan table move}),
			so that the simulation may resume from there.
			If at some point~$\hp=0$ and no position has been found and selected,
			then the machine aborts (\cref{l:nsimult/scan table abort}).
	\end{enumerate}
	By construction, $\mach{C}_i$ simulates~$\mach{A}_{/R}$ on relative position less than~$n^2+i$.
	Next, its transitions are determined from the only information of~$\var{\p_{curr}}$ that ranges over~$\Q$, $\hp$ and the scanned symbol.
	Hence, $\mach{C}_i$ has~$n$ states (not counting the \hp\=/component).%
\end{proof}

Denote by \defem{$\mach{C}$} the disjoint union of~$\mach{C}_i$'s over~$i$,
that is, the \twnfa implementing~\cref{proc:nsimult}.
A computation of~$\mach{C}$ may fail for distinct reasons.
First, it may follow a computational path of~$\mach{A}_{/R}$ on~$\pi_1(y)$
that gets stuck in some configuration with the head positioned on some symbol of~$\lend\pi_1(y)$.
This includes entering a configuration in which~$\hp=n^2-1$ (last position of~$x$)
with~$\var{\p_{curr}}$ storing some state~$\p$
for which~$R(\p)=\emptyset$.
Second, from such a point, even if~$R(\p)\neq\emptyset$,
$\mach{C}$ may choose none of the~$\q$ of this set (by choosing~$b=\False$ on \cref{l:nsimult/scan table choice})
and finally end in abortion (on \cref{l:nsimult/scan table abort}).
Third, it may follow a computational path of~$\mach{A}_{/R}$ on~$\pi_1(y)$ that enters a loop.
In such a case the simulation might loop forever.
This last case is more problematic, as it cannot be easily detected.
However, the following remark shows that,
with a polynomial increase in size,
it is possible to avoid such loops,
while keeping the main property of the automaton.
\begin{remark}[About haltingness of~$\mach{C}$]
	\label{res:nsimult halting}
	Since, for every~$i$,~$\mach{C}_i$ works within the current window of length~$2n^2$ and has~$n$ states,
	every computational path of~$\mach{C}$
	with more than~$2n^3-1$ steps
	admits a loop (\ie, visits twice the same configuration).
	Such a loop could have been cut,
	hence resulting in a shorter computational path
	with same starting and ending configurations.
	By adding a clock component to the finite control,
	as a variable ranging over~$\interv{2n^3}$,
	one may restrict computational paths of~$\mach{C}$
	to have length at most~$2n^3-1$.
	This preserves all loop-free computational paths
	(and thus the main property of~$\mach{C}$)
	and ensures haltingness.
	Moreover, the size of the resulting $\hp$\=/aware halting variant of~$\mach{C}$
	is still polynomial in~$n$, namely, the number of states is in~$\bigO{n^5}$.%
\end{remark}

In order to update the \Ltables according to \cref{res:update ltables},
we need to consider other relations, \eg~$\lS[\mach{A}_{/R}]*{z\sigma}{j}$ for~$j\in\interv{n^2}$.
The automaton~$\mach{C}$ can be easily adapted so that the resulting \twnfa
is able to find any computational path of~$\mach{A}_{/R}$
witnessing the membership of a pair~$(\p,\q)$ to such relations.
\begin{lemma}
	Let~$i\in\interv{n^2}$.
	There exists an $\hp$\=/aware \twnfas~$\mach{S}_i$ with state set~$\Q\times\interv{n^2}$
	such that,
	for every~$x\in\set{\lend}\cup\analphab^{n^2}$,
	$y\in\analphab^{i}$,
	$\sigma\in\analphab\cup\set{\rend}$,
	$\p,\q\in\Q$,
	and
	$j\in\interv{n^2}$:
	\begin{align*}
		\cfg{xy}{\hpQ{(\p,j)}{n^2+i}}{\sigma}	&	\yield*[\mach{S}_i]	\cfg{xy}{\hpQ{(\q,0)}{n^2+i}}{\sigma}
		&	\iff &&
		(\p,\q)\in\lS[\mach{A}_{/R}]{\pi_1(y\sigma)}{j}
	\end{align*}
	where~$R$ equals~$\lt[A]{\ew}$ if~$x=\lend$ and~$\decode{\pi_2(x)}$ otherwise.
	Furthermore,
	$\mach{S}_i$ forbids right moves from configurations in which~$\hp=n^2+i$,
	and has no transition outgoing configurations in which~$\hp>n^2+i$ or the state is~$(\q,0)$ for some~$\q$;
	in particular the above configuration~$\cfg{xy}{(\q,0)}{\sigma}$ is halting.
\end{lemma}
\begin{proof}
	On input~$xy\sigma$,
	based on \cref{rk:T1 from t},
	an \hp\=/aware \twnfa witnesses the membership of a pair to~$\lT[\mach{A}_{/R}]*{\pi_1(y\sigma)}{1}$ as follows.
	Starting from the relative position~$n^2+i$ (thus scanning~$\sigma$),
	it simulates one left-move of~$\mach{A}_{/R}$ over~$\pi_1(\sigma)$,
	and then, using~\nsimult implemented by the automaton~$\mach{C}_i$ from \cref{res:nsimult},
	it simulates an \Lseg of~$\mach{A}_{/R}$ on~$\pi_1(y)$,
	thus ending in relative position~$\length{y\sigma}$.
	By repeating this process up to~$j$ times
	(using a counter ranging over~$\interv{j}$),
	it can witness membership of a pair to~$\lS[\mach{A}_{/R}]*{\pi_1(y\sigma)}{j}$.
	Only~$\bigO{n^3}$ states are sufficient to implement this process,
	namely for storing the counter ranging over~$\interv{j}\subseteq\interv{n^2}$,
	and the state of~$\mach{A}_{/R}$ in the simulated computation.
\end{proof}
We let~\defem{$\mach{S}$} be the disjoint union of~$\mach{S}_i$'s over~$i$.
In subsequent procedures, the call to~$\mach{S}$ is referred to as the procedure~\defem{\nsimulS}
which takes two parameters, namely the starting state~$\p$ and the value of~$j$,
and eventually returns a state~$\q\in\lS[A]{z\sigma}{j}(\p)$ when called from head position~$\length{z\sigma}$.

Let~$Y$ be~$\lt[A]{v\tau}$ (\resp~$\lS[A]{v\tau}{j}$ for some~$j$), and~$m$ denote the size of~$Y$.
Equipped with the procedure~\nsimult (\resp~\nsimulS),
a \twnfa can enumerate the elements of~$Y$ as soon as~$m$ is known.
Indeed,
in a similar way as \cref{proc:enumQX} enumerates the elements of~$\QX[A]{v}$,
it is possible to repeat~$m$ calls to~$\nsimult$ (\resp~$\nsimulS$)
and check that at each iteration but the first,
the found pair is larger than the preceding one.
The so-described enumeration procedure is named \defem{$\enumt$} (\resp \defem{$\enumS$}{}),
and can be implemented by a $\hp$\=/aware \twnfa with~$\bigO{n^8}$ (\resp~$\bigO{n^{10}}$) many states.
In particular,
\enumt has to store the value of the variable \var{k} 
for which it uses a state component of size at most~$n^2$,
and~$n$-size state components for the values of~$\p_{prev}$, $\q_{prev}$, $\p_{next}$, and $\q_{next}$.
Notice that the variable~$\q_{next}$ is not used during the execution of~\nsimult,
and then overwritten with the value returned by~$\nsimult$,
hence it is not needed to be stored in the finite control while executing~\nsimult.
Therefore, that state component can be used to store the value of the internal state variable~$\var{\p_{curr}}$ of~$\nsimult$.
In addition,
\nsimult uses a state component of size~$n^2$ for the variable~$i$.
So,
summing up the sizes of the state components we end up with~$\bigO{n^8}$ states.
On the other hand,
\enumS uses the same states as \enumt,
plus a state component of size~$n^2$ for the variable~$j$.
\begin{figure}
	\begin{multicols}{2}
		\begin{procedure}[H]
			\caption{%
				()\space%
				\enumt{$m$}
			}
			$(\var{\p_{prev}},\var{\q_{prev}})\gets (-1,-1)$\;
			\For{$\var{k}\gets 1$ \KwTo~$m$}{%
				choose~$\var{\p_{next}}$ in~$\Q$\;
				$\var{\q_{next}}\gets\nsimult(\var{\p_{next}})$\;
				\lIf{$(\var{\p_{next}},\var{\q_{next}})\leq(\var{\p_{prev}},\var{\q_{prev}})$}{\Abort}
				$(\var{\p_{prev}},\var{\q_{prev}})\gets (\var{\p_{next}},\var{\q_{next}})$\;
				\KwYield $(\var{\p_{next},\var{\q_{next}}})$\;
			}
		\end{procedure}
		\begin{procedure}[H]
			\caption{%
				()\space%
				\enumS{$m$, $j$}
			}
			$(\var{\p_{prev}},\var{\q_{prev}})\gets (-1,-1)$\;
			\For{$\var{k}\gets 1$ \KwTo~$m$}{%
				choose~$\var{\p_{next}}$ in~$\Q$\;
				$\var{\q_{next}}\gets\nsimulS(\var{\p_{next}}, $j$)$\;
				\lIf{$(\var{\p_{next}},\var{\q_{next}})\leq(\var{\p_{prev}},\var{\q_{prev}})$}{\Abort}
				$(\var{\p_{prev}},\var{\q_{prev}})\gets (\var{\p_{next}},\var{\q_{next}})$\;
				\KwYield $(\var{\p_{next},\var{\q_{next}}})$\;
			}
		\end{procedure}
	\end{multicols}
\end{figure}

\paragraph{Checking the annotation-track contents.}
\label{sec:checktable}
The above-presented procedures~\enumt and~\enumS,
respectively,
allow to enumerate the elements of~$\lt[\mach{A}_{/R}]*{\pi_1(y)}$ and~$\lS[\mach{A}_{/R}]*{\pi_1(y)}{j}$ for some~$y$ and~$j$,
where~$R$ is encoded over the annotation track on the left side of the current window
(\ie, on the cells at relative positions from~$0$ to~$n^2-1$)
which is scanned during the inner calls to~\nsimult and~\nsimulS.
In order to get membership to~$\lt[A]{v}$ and~$\lS[A]{v}{j}$
where~$v$ is the whole input-track contents to the left of the current position,
we should ensure the correctness of~$R$.
More precisely,
if~$\lend w\rend$ is factorized as~$\lend w'xyw''\rend$
for some~$w'$ whose length is a multiple of~$n^2$,
some~$x$ of length~$n^2$ such that~$\decode{\pi_2(x)}=R=\lt[A]{\pi_1(w'x)}$
some~$y$ of length~$n^2$,
and some~$w''$,
then~$\mach{B}$ should check that~$\decode{\pi_2(y)}=\lt[A]{\pi_1(w'xy)}=\lt[\mach{A}_{/R}]*{\pi_1(y)}$
before entering the first position of~$w''\rend$.
Since, by induction hypothesis,
the automaton will know~$\card{\lt[A]{\pi_1(w'xy)}}$ at that time
(meaning that~$\card{\lt[A]{\pi_1(w'xy)}}$ will be stored in the finite control),
it can enter a special mode to perform the check.
The same idea as in the one-way case (see \cref{proc:checkAnnot})
can be used,
namely,
the automaton first checks that the annotation track contains exactly~$\card{\lt[A]{\pi_1(w'xy)}}$ many~$1$'s
and then, using the procedure~\enumt,
it checks that the elements of~$\lt[A]{\pi_1(w'xy)}$ are exactly those annotated by~$1$ in the annotation track.
This can be implemented by using in~$\bigO{n^8}$ many states.

Although useless for the simulation,
in order to fulfill the requirement of uniqueness of~$\annot{w}$,
when hitting the right endmarker
$\mach{B}$ should check that the last~$r$ input cells were annotated by~$0$,
where~$r=\length{w}\mod n^2$
(\cf \cref{def:annot}).
This is easily done using $\hp$\=/awareness.

\subsubsection{Inductively computing the size of \Ltables.}
\label{sec:inductive counting}
We now explain how our \twnfa~$\mach{B}$ inductively computes~$\card{\lt[A]{\pi_1(v)}}$ for each prefix~$v$ of~$w$.
More precisely,
we inductively ensure the following invariant for every prefix~$v$ of~$w$:
\begin{equation}
	\label{invariant}
	\tag{$I_v$}
	\begin{gathered}
		\text{%
			when~$\mach{B}$ enters position~$\length{v}+1$ for the first time,
			$\card{\lt[A]{v}}$ is stored in its finite control,%
		}%
		\\
		\text{%
			and~$\pi_2(v')=\encode{\lt[A]{\pi_1(v')}}$ for~$v'$ the maximal prefix of~$v$ such that~$\length{v'}=0\mod n^2$.%
		}
	\end{gathered}
\end{equation}
Initially, for~$v=\ew$, $\card{\lt[A]{v}}=\card{\lt[A]{\ew}}$ is a precomputed constant, encoded in the initial state of~$\mach{B}$.
Let~$v\sigma$ be a prefix of~$w$ with~$\length{\sigma}=1$.
Assume that~$\card{\lt[A]{v}}$ is known (\ie, stored in the finite control of~$\mach{B}$)
and the head is on position~$\length{v\sigma}$, scanning~$\tau$ such that~$\pi_1(\tau)=\sigma$.
In order to compute~$\card{\lt[A]{v\sigma}}$,
$\mach{B}$ follows the following steps:
\begin{enumerate}[label={Step~\arabic*.},font={\color{lipicsGray}\sffamily\bfseries\upshape\mathversion{bold}},left={.5em},ref=\arabic*]
	\item\label{it:S1 from t}
		It computes~$\card{\lS[A]{v\sigma}{1}}$ from~$\card{\lt[A]{v}}$ (\cref{proc:S1 from t});
	\item\label{it:S* from S1}
		It then inductively computes~$\card{\lS[A]{v\sigma}{*}}$ from~$\card{\lS[A]{v\sigma}{1}}$ (\cref{proc:S* from S1}, using \cref{proc:Snext from Sprev});
	\item\label{it:t from S*}
		It finally computes~$\card{\lt[A]{v\sigma}}$ from~$\card{\lS[A]{v\sigma}{*}}$ (\cref{proc:t from S*}).
\end{enumerate}
Each of these steps is performed through a computational path of~$\mach{B}$
that starts from position~$\length{v\sigma}$, ends in position~$\length{v\sigma}$,
and visits only position~$h\leq\length{v\sigma}$ in the meantime.

\xref[Steps]{it:S1 from t} and~\ref{it:t from S*} are the simplest ones,
as they follow from~\cref{rk:T1 from t,res:t from S*}, respectively.
Indeed, for \hyperref[it:S1 from t]{Step~\ref*{it:S1 from t}},
\mach{B} can detect whether a given pair~$(\p,\q)$ belongs to~$\lt[A]{v}$ given~$\card{\lt[A]{v}}$ using~$\membert$.
Thus, it can detect whether a pair belongs to~$\lS[A]{v\sigma}{1}$ based on \cref{rk:T1 from t},
and, by trying all pairs and counting those that belong to~$\lS[A]{v\sigma}{1}$,
it can compute~$\card{\lS[A]{v\sigma}{1}}$ (\cref{proc:S1 from t}).
Similarly for \hyperref[it:t from S*]{Step~\ref*{it:t from S*}},
$\mach{B}$ can detect whether a given pair~$(\p,\q)$ belongs to~$\lS[A]{v\sigma}{*}$ given~$\card{\lS[A]{v\sigma}{*}}$
using~$\memberS$ with~$j=n^2$
(so that~$\lS[A]{v\sigma}{j}=\lS[A]{v\sigma}{*}$ by \cref{res:iterated ltables bound}).
Thus, it can detect whether a pair belongs to~$\lt[A]{v\sigma}$ based on~\cref{res:t from S*},
and by trying all pairs and counting those that belong to~$\lt[A]{v\sigma}$,
it can compute~$\card{\lt[A]{v\sigma}}$ (\cref{proc:t from S*}).
The resulting procedures~\getSonefromt and~\gettfromSstar
can be implemented using a polynomial number of states.
Indeed, \membert\ (\resp \memberS) is implemented using~$f(n)$ states for some polynomial~$f$,
and thus~\getSonefromt (\resp \gettfromSstar) can be implemented using~$\bigO{n^5f(n)}$ states
(the $\bigO{n^5}$\=/size additional state component allows to store~$\p$, $\q$, $\r$, and~$m$).
\begin{figure}
	\begin{multicols}{2}
		\begin{procedure}[H]
			\caption{%
				()\space%
				\getSonefromt{$\var{m}$}
				\label{proc:S1 from t}
			}
			$\var{m_{next}}\gets n$\tcp*[f]{$n$ for~$\set{(\p,\p)}[\p\in\Q]$}\;
			\ForEach{$(\p,\q)\in\set{(\p,\q)\in\Q^2}[\p\neq\q]$}{%
				\ForEach{$(\q[s],\r)\in\enumt{$m$}$}{%
					\If{%
						$\r=\q$ \And $(\q[s],\lmove)\in\delta(\p,\pi_1(\read{}))$%
					}{%
						$\var{m_{next}}\gets\var{m_{next}}+1$\;
						\Break\;
					}
				}
			}
			\Return $\var{m_{next}}$;
		\end{procedure}
		\begin{procedure}[H]
			\caption{%
				()\space%
				\gettfromSstar{$\var{m}$}%
				\label{proc:t from S*}
			}
			$\var{m_{next}}\gets0$\;
			\ForEach{$(\p,\q)\in\Q^2$}{%
				\ForEach{$(\q[s],\r)\in\enumS{$m$, $n^2$}$}{%
					\If{%
						$\q[s]=\p$ \And $(\q,\rmove)\in\delta(\r,\pi_1(\read{}))$
					}{%
						$\var{m_{next}}\gets\var{m_{next}}+1$\;
						\Break\;
					}
				}
			}
			\Return $\var{m_{next}}$;
		\end{procedure}
	\end{multicols}
	\vskip-1.75\baselineskip
\end{figure}

It thus remains to explain \hyperref[it:S* from S1]{Step~\ref*{it:S* from S1}},
which follows the same idea but with an inner induction.
Indeed, one way of computing~$\card{\lS[A]{v\sigma}{*}}=\card{\lS[A]{v\sigma}{n^2}}$ from~$\card{\lS[A]{v\sigma}{1}}$
is to successively compute~$\card{\lS[A]{v\sigma}{j}}$ for ${j=1,\ldots,n^2}$.
However, we follow a cheaper and simpler big-step strategy,
where we compute these cardinalities only for successive powers of~$2$,
using the following observation:
\begin{align}
	(\p,\q)\in\lS[A]{v\sigma}{2j}
	&&\iff&&
	\exists\r\in\Q\mathinner{:}\;
	(\p,\r)\in\lS[A]{v\sigma}{j}
	\mit{and}
	(\r,\q)\in\lS[A]{v\sigma}{j}
	&&
	\text{for all~$j\geq0$}
	\text.
\end{align}
Hence, since knowing~$\card{\lS[A]{v\sigma}{j}}$ allows
to detect which pairs of states belong to~$\lS[A]{v\sigma}{j}$ using $\memberS$,
our automaton~$\mach{B}$ is able to compute~$\card{\lS[A]{v\sigma}{2j}}$ from~$\card{\lS[A]{v\sigma}{j}}$.
Just as~\gettfromSstar,
the resulting procedure~\getSnextfromSprev
can be implemented using a polynomial number of states only (\cref{proc:Snext from Sprev}).
By successively computing such cardinalities for~${j=1,\ldots,2\ceil{\log(n)}}$,
we obtain~$\card{\lS[A]{v\sigma}{2^{2\ceil{\log(n)}}}}$
which is equal to~$\card{\lS[A]{v\sigma}{*}}$ by \cref{res:iterated ltables bound}.
The resulting procedure \getSstarfromSone (\cref{proc:S* from S1})
can be implemented using a polynomial number of states in~$n$ only.
\begin{figure}
	\begin{multicols}{2}
		\begin{procedure}[H]
			\caption{%
				()\space%
				\getSnextfromSprev{$\var{m}$, $j$}%
				\label{proc:Snext from Sprev}
			}
			$\var{m_{next}}\gets 0$\;
			\ForEach{$(\p,\q)\in\Q^2$}{%
				\ForEach{$\r\in\Q$}{%
					\If{$\memberS{\p, \r, $\var{m}$, $j$}$ \And $\memberS{\r, \q, $\var{m}$, $j$}$}{%
						$\var{m_{next}}\gets\var{m_{next}}+1$\;
						\Break\;
					}
				}
			}
			\Return $\var{m_{next}}$\;
		\end{procedure}
		\begin{procedure}[H]
			\caption{%
				()\space%
				\getSstarfromSone{$\var{m}$}%
				\label{proc:S* from S1}
			}
			$\var{m_{next}}\gets\var{m}$\;
			\For{$j\gets1$ \KwTo~$2\ceil{\log(n)}$}{%
				$\var{m_{next}}\gets\getSnextfromSprev{$\var{m_{next}}$, $2^j$}$\;
			}
			\Return $\var{m_{next}}$\;
		\end{procedure}
	\end{multicols}
	\vskip-1.75\baselineskip
\end{figure}

\subsubsection{Gathering the mechanisms: the automaton~$\mach{B}$.}
\label{sec:gathering pieces}
We are now ready to prove our main theorem,
by concluding the presentation of~$\mach{B}$,
whose high-level behavior is described by \cref{proc:mainB}.
\begin{procedure}[tb]
	\caption{%
		()\space%
		\mainB{}%
		\label{proc:mainB}
	}
	$\var{m}\gets\card{\lt[A]{\ew}}$\;
	\While(\label{l:mainB/loop/begin}){$\read{}\neq\rend$}{%
		$\var{m}\gets\getSonefromt{\var{m}}$
		\label{l:mainB/loop/S1 from t}
		\;
		$\var{m}\gets\getSstarfromSone{\var{m}}$
		\label{l:mainB/loop/S* from S1}
		\;
		$\var{m}\gets\gettfromSstar{\var{m}}$
		\label{l:mainB/loop/t from S*}
		\;
		\lIf(%
		\label{l:mainB/loop/checktable}%
		){$\hp=2n^2-1$}{
			\checktable{$\var{m}$}
		}
		move the head to the right
		\label{l:mainB/loop/head shift}
		\label{l:mainB/loop/end}
		\;
	}
	\lIf(\label{l:mainB/final/checkannot}){the suffix of length~$\hp-n^2-1$ of~$\pi_2(w)$ is not in~$0^*$}{\Abort}
	$\var{m}\gets\getSonefromt{\var{m}}$
	\label{l:mainB/final/S1 from t}
	\;
	$\var{m}\gets\getSstarfromSone{\var{m}}$
	\label{l:mainB/final/S* from S1}
	\;
	\leIf(\label{l:mainB/final/accept}\label{l:mainB/final/reject}){$\enumS{$\var{m}$, $n^2$}$ outputs $(\qrestart, \qfin)$}%
	{\Return \True}%
	{\Return \False}%
\end{procedure}
\begin{lemma}
	\label{res:2nfa checking annotation and testing membership}
	Let~$\mach{A}$ be an~$n$\=/state \twnfa over~$\ialphab$,
	and let~$\analphab=\ialphab\times\set{0,1}$.
	Then there exist a function~$\annotname:\ialphab*\to\analphab*$,
	and a \twnfa~$\mach{B}$ of polynomial size in~$n$%
	\footnote{%
		In order to avoid technicalities,
		and because the polynomial order of our simulation is what matters here,
		we do not give a precise polynomial upper bound.
		A rough evaluation could show that~$\bigO{n^{17}\log(n)}$ states are sufficient.%
	}
	over~$\analphab$,
	with two distinguished halting states~$\qacc$ and~$\qrej$
	such that, on input~$x\in\analphab*$:
	\begin{itemize}
		\item $\mach{B}$ admits an initial computational path halting in~$\qacc$ \ifof~$x=\annotname(\pi_1(x))$ and~$\pi_1(x)\in\langof{A}$;
		\item $\mach{B}$ admits an initial computational path halting in~$\qrej$ \ifof~$x=\annotname(\pi_1(x))$ and~$\pi_1(x)\notin\langof{A}$.
	\end{itemize}
\end{lemma}
\begin{proof}
	The mapping~$\annotname$ is defined in~\cref{def:annot},
	while the \twnfa~$\mach{B}$ implements~\cref{proc:mainB}
	using and maintaining the~$\hp$ variable in its finite control,
	as explained in \cref{sec:sliding window}.
	We first show the correctness of~$\mach{B}$,
	based on the preliminary work,
	and then argue that its has polynomial size in~$n$.

	The behavior of~$\mach{B}$
	can be decomposed into two phases,
	one main loop for visiting all tape cells until hitting the right endmarker (\crefrange{l:mainB/loop/begin}{l:mainB/loop/end})
	while iteratively computing~$\card{\lt[A]{v}}$ for the corresponding prefix~$v$,
	followed by a final phase deciding the acceptance or rejection of the input.
	In any phase, abortion of the computation, due to wrong choices or incorrect annotations are possible.

	During the first phase,
	we ensure the invariant~\eqref{invariant}
	where~$v$ is the input-track contents to the left of the head position,
	at the loop entrance.
	This prefix~$v$ is extended, symbol by symbol, at each iteration of the loop (see \cref{l:mainB/loop/head shift}).
	Indeed, knowing~$\card{\lt[A]{v}}$ at the beginning of the loop with the head scanning~$\tau\neq\rend$ with~$\pi_1(\tau)=\sigma$,
	the machine computes~$\card{\lt[A]{v\sigma}}$
	by following the three steps that were presented in \cref{sec:inductive counting}
	(\crefrange{l:mainB/loop/S1 from t}{l:mainB/loop/t from S*}).
	In order for the simulation to work properly,
	$\mach{B}$ checks the annotation track contents
	using the procedure~\checktable
	each time a factor of length~$n^2$ have been completed
	(\cref{l:mainB/loop/checktable}).

	The second phase is entered when the head has reached the right endmarker.
	In order to ensure unicity of the annotation,
	it checks that the annotation-track contents ends with~$0^r$,
	where~$r=\length{w}\mod n^2$ is known as~$r=\hp-n^2-1$ (\cref{l:mainB/final/checkannot}).
	Also, at that time,
	the state of~$\mach{B}$ contains the information of~$\card{\lt[A]{w}}$.
	Hence, in a similar way as done during the loop,
	the machine can compute~$\card{\lS[A]{w\rend}{*}}$ (\crefrange{l:mainB/final/S1 from t}{l:mainB/final/S* from S1}).
	It then decide acceptance by testing whether the pair~$(\qrestart,\qfin)$ belongs to~$\lS[A]{w\rend}{*}$ or not according to~\cref{res:ltables for recognition},
	using~$\enumS$ with~$j=n^2$ (\cref{l:mainB/final/accept}).

	The number of states of~$\mach{B}$ is polynomial in~$n$.
	This can easily be seen as the given procedure and sub-procedures
	use finitely many variables (including~\hp),
	each ranging over at most~$n^2$ values.
	Indeed a state of~$\mach{B}$ roughly consists in a valuation of these variables,
	together with a mode specifying at which point (which procedure and which line) the simulation is.
	A rough analysis could show that~$\bigO{n^{17}\log(n)}$ states are sufficient
	(hopefully a finer analysis and/or design could decrease this exponent,
	possibly using further annotation symbols).
\end{proof}

Our main theorem then follows.
\begin{theorem}
	\label{thm:2nfa to sv1-la}
	Every $n$\=/state \twnfa
	has an equivalent self-verifying \twnfacg or \la
	with a polynomial number of states in~$n$,
	and~$2$ annotation symbols.
\end{theorem}

\end{document}

\section{Conclusion}
\label{sec:conclusion}

We investigated the descriptional complexity of the complementation of \twnfas.
Our results show that,
if we relax the target device so that,
in addition to the usual nondeterminism
it enjoys \emph{common guess}
(\ie, it works on nondeterministically annotated inputs),
then a polynomial cost can be met.
Working on \cref{res:nsimult halting} or applying the results from~\cite{GP19},
the resulting \twnfacg can furthermore be made halting.
A natural improvement of our result consists in a finer analysis of the upper-bounding polynomial,
or --more promising-- in the design of an alternative cheaper construction, possibly using more annotation symbols.

Another extension could be to consider more restricted forms of~\twnfacgs as target devices.
In particular, \twdfacgs are of particular interest.
In contrast with~\twnfacgs,
their nondeterminism is limited to the annotation phase only.
This limitation already allows to derive similar lower bounds
for their simulation by \twnfas, \ownfas, \owdfas, or deterministic~\las (\dlas)
as those obtained for the simulation of general~\las~\cite{GPT25}.
A natural question is whether common guess is sufficient for capturing the usual form of nondeterminism
up-to a polynomial size increase.
This reduces to the question of the size cost of the conversion of~\twnfas into equivalent~\twdfacgs,
a weakening of~\cite[Problem~4]{Pig19}; see~\cref{fig:map}.
Hence, a future line of research consists in the investigation of these costs,
as well as the related problem of the cost of complementing~\twnfas or~\ownfas by~\twdfacgs.

\bibliography{references}
\cleardoublepage
\end{document}